\documentclass[aps,prl,twocolumn,groupedaddress,superscriptaddress,showpacs,preprintnumbers,amsmath,amssymb,floatfix]{revtex4}

\usepackage{graphicx}
\usepackage{stackrel}

\newlength{\onefig}
\newlength{\twofig}
\newcommand{\ve}[1]{\mathbf{#1}}

\usepackage[usenames,dvipsnames]{color}

\begin{document}

\title{Clustering, conductor-insulator transition and phase
  separation of an ultrasoft model of electrolytes}

\author{Daniele Coslovich}
\affiliation{Laboratoire Charles Coulomb, Universit\'e Montpellier 2 et CNRS, Montpellier, France}
\author{Jean-Pierre Hansen}
\affiliation{Departement of Chemistry, University of Cambridge, Cambridge, United Kingdom} 
\author{Gerhard Kahl}
\affiliation{Institut f\"ur Theoretische Physik and CMS, Technische Universit\"at Wien, Vienna, Austria} 

\date{\today}

\begin{abstract}
  We investigate the clustering and phase separation of a model of
  ultrasoft, oppositely charged macroions by a combination of Monte
  Carlo and Molecular Dynamics simulations. Static and dynamic
  diagnostics, including the dielectric permittivity and the electric
  conductivity of the model, show that ion pairing induces a sharp
  conductor-insulator transition at low temperatures and densities,
  which impacts the separation into dilute and concentrated phases
  below a critical temperature. Preliminary evidence is presented for
  a possible tricritical nature of the phase diagram of the model.
\end{abstract}

% \pacs{82.70.Dd, 64.75.Xc, 66.10.Ed, 64.60.Kw}

\maketitle

Clustering of oppositely charged ions is a common
mechanism~\cite{bjerrum__1926,stillinger_ion-pair_1968} which strongly
influences the thermodynamic and transport properties of electrolytes,
in particular the phase separation into dilute and concentrated ionic
solutions predicted to occur at low
temperatures~\cite{panagiotopoulos_simulations_2005}. This phase
separation (akin to vapour-liquid coexistence) has been investigated
in considerable detail both theoretically and by computer simulations
in the context of the primitive model of electrolytes, consisting of
oppositely charged hard spheres immersed in a dielectric continuum
(implicit solvent), with most of the published work focusing on the
``restricted'' version of the model (RPM) featuring equal anion and
cation
diameters~\cite{panagiotopoulos_simulations_2005,stell_critical_1976,levin_criticality_1996,orkoulas_phase_1999,caillol_critical_2002}. The
RPM is a reasonable model for strong, monovalent electrolytes, like
aqueous solutions of NaCl. The formation of long-lived dipolar pairs
considerably affects the coexistence
curve~\cite{levin_criticality_1996,romero-enrique_dipolar_2002}, and
severely limits the ergodicity of traditional Monte Carlo (MC) or
Molecular Dynamics (MD) simulations at low temperatures~\cite{valeriani_computer_2010}.

Recently the attention has shifted to the rich phase behaviour of ``colloidal electrolytes'' where
the anions and cations are highly charged, hard
colloidal~\cite{leunissen_ionic_2005} or nanometric
particles~\cite{ryden_monte_2005}, usually
in the presence of added salt~\cite{hynninen_cuau_2006,caballero_complete_2007,sanz_gel_2008}. 
In this Letter, we extend the RPM to a broader class of soft matter
systems by introducing an ``ultrasoft'' restricted primitive model (URPM), where
macroions are assumed to be penetrable charged particles, i.e., the
hard cores of charged colloids are replaced by bounded interactions at
short interionic distances. Ultrasoft core representations of the
effective interaction between the centres of mass (CM) of polymer
coils have proved very
successful in describing dilute and semi-dilute polymer
solutions~\cite{bolhuis_accurate_2001,pierleoni_multiscale_2006}. Our model generalises such a
representation to solutions of oppositely charged polyelectrolytes
chains~\cite{phillip_98,buchhammer_03}. %such as unfolded proteins~\cite{ritort_single-molecule_2006}, charged dendrimers~\cite{huimann_effects_2010}, or micelles~\cite{wu_interaction_2000}. 
We explore, through MC and MD simulations, the subtle interplay between the clustering effects
associated with interpenetrating soft core
particles~\cite{likos_do_2007} and the long-range Coulombic
interactions, and its influence on phase separation.
Our simulation results for the URPM reveal a non-trivial topology of the
phase diagram of the model 
% (especially when compared to that of the RPM~\cite{valeriani_ion_2010})
and suggest a strong link between phase separation and a purely classical
conductor-insulator (CI) transition at low temperatures, reminiscent of
the behaviour of liquid metals~\cite{hensel_liquid-vapour_1995}.

The URPM is a system of $n_+$ cations of total charge $+Q$ and $n_-$
anions of charge $-Q$ per unit volume moving in a dielectric continuum
of relative permittivity $\epsilon'$; global charge neutrality implies
$n_+ = n_- = n$. The charge around the CM of each macroion follows a
Gaussian distribution $Q_\alpha \rho_\alpha(s) = Q_\alpha \exp[-s^2/2
\sigma^2] \left( 2 \pi \sigma^2 \right)^{-3/2}$ ($\alpha$ = + or -)
where $s$ is the distance from the CM. The resulting pair potentials
as functions of the distance $r$ between the CM's of two ions are:
\begin{eqnarray}
\nonumber
v_{\alpha \beta}(r) & = & \frac{Q_\alpha Q_\beta}{\epsilon' r} {\rm erf}(r/2 \sigma) \\
\label{eqn:1b}
 & \stackrel[r \to 0]{}{\approx} & \pm u_0 \left[ 1 - \frac{r^2}{12 \sigma^2} + {\cal O}(r^4) \right]
\end{eqnarray}
where $-u_0 = - Q^2/\sqrt{\pi} \epsilon' \sigma$ is the energy of a
fully overlapping anion/cation pair ($r=0$). For r $\gtrsim \sigma$,
the pair potentials go over to the Coulombic interaction between point
charges. A related model was used previously to investigated a very
different system, namely a semi-classical Hydrogen
plasma under astrophysical conditions of high temperatures and
densities~\cite{hansen_microscopic_1981}. In the following, we will use 
$\sqrt{2}\sigma$, $u_0$, and $\sqrt{2 m\sigma^2/u_0}$ as units
of length, energy and time, respectively.
We will report MC and MD simulations results over a broad range
of total density $\rho=2n$ and temperature $T$, obtained for samples of $N = N_+ + N_- =
1000$ anions and cations under periodic boundary conditions, employing the Ewald
summation technique.
Comparison of different
simulations methods and different thermal histories provide evidence
of proper equilibration of our samples over the investigated range
of state parameters.

The classical ground state energy ($T = 0$) is $U = - N u_0$, which
is extensive, ensuring the existence of a proper thermodynamic
limit~\cite{ruelle_statistical_1999}. In the high temperature limit ($T \agt 0.5$) we found that the model
is accurately described by the random phase approximation
(RPA), which reduces to the familiar Debye-H\"uckel theory for point
ions and has been shown to be very accurate also for other soft core
models~\cite{bolhuis_accurate_2001,pierleoni_multiscale_2006,likos_do_2007}
at high densities.

\begin{figure}[tbp]
\includegraphics[width=\onefig]{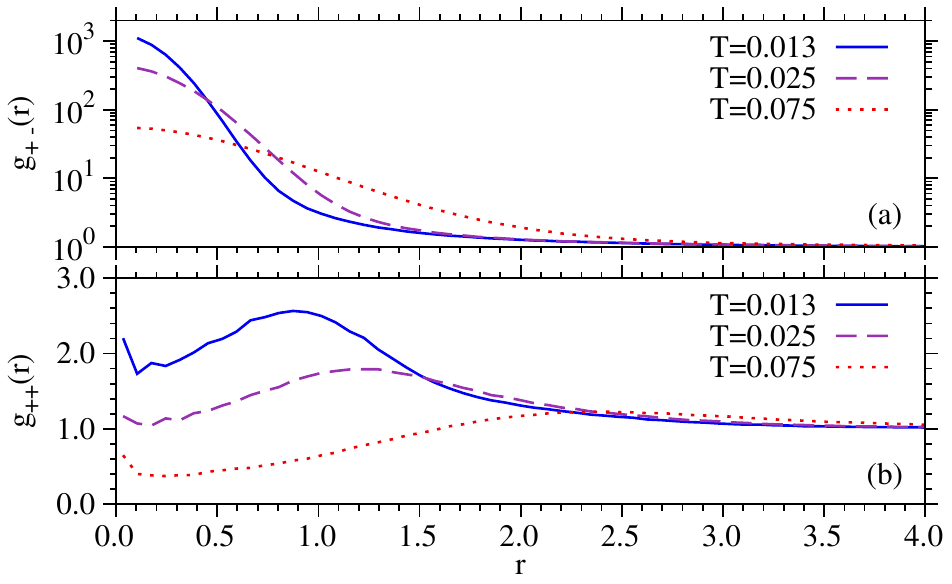}
\caption{\label{fig:1} Radial distribution functions for selected
  temperatures (as indicated) along the isochore $\rho=0.01$: (a)
  $g_{+-}(r)$, (b) $g_{++}(r)$.}
\end{figure}

An interesting phenomenology appears at sufficiently low
densities and temperatures, where the system shows clear signatures of
clustering. To quantify this phenomenon we first inspect the equilibrium
pair structure, which is characterised by the two radial distribution
function (RDF) $g_{++}(r) = g_{--}(r)$ and $g_{+-}(r)$. MD data are shown in
Fig.~\ref{fig:1}, along the isochore $\rho = 0.01$, representative of the behaviour of the system at low density. At low $T$ the
sharp rise of $g_{+-}(r)$ as $r \to 0$ points to strong anion/cation
clustering, as evident in snapshots of ion configurations. Furthermore, clustering at
low $T$ is also indirectly evident from $g_{++}(r)$, which appears
to imply an attraction between equally charged ions [$g_{++}(r
\lesssim \sigma) > 1$]---a fact that can only be rationalised by the formation
of tight neutral pairs or higher order clusters.

\begin{figure}[t]
\includegraphics*[width=\onefig]{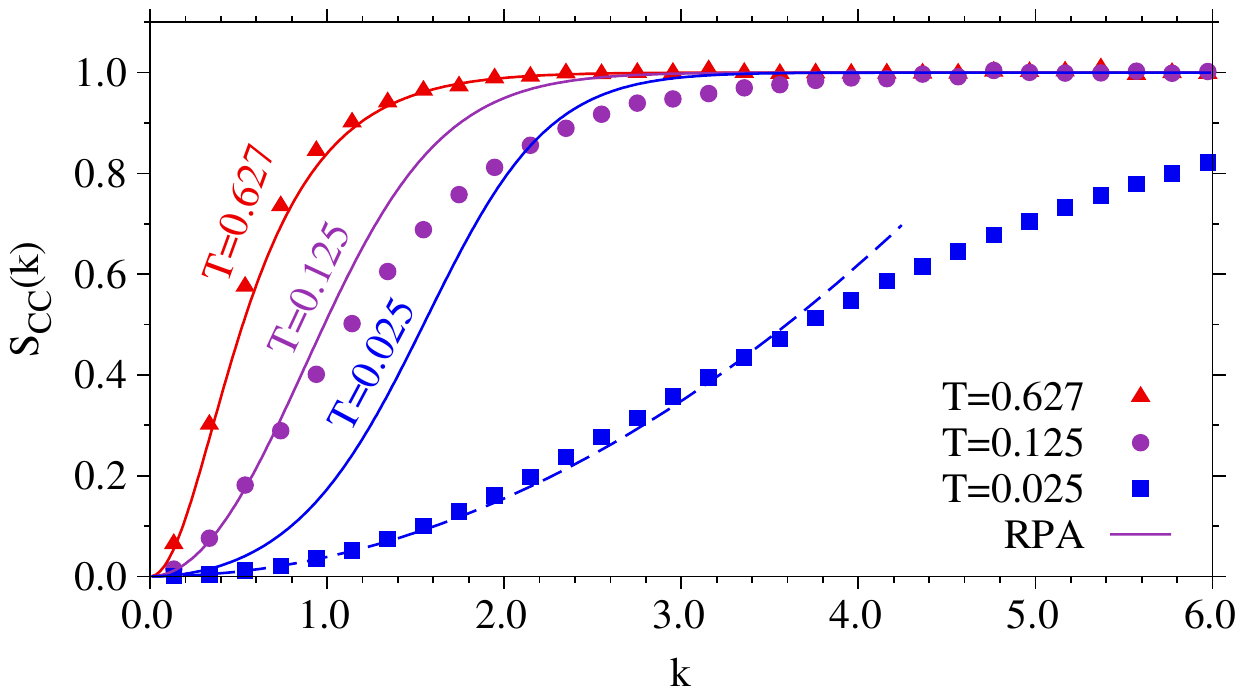}
\caption{\label{fig:5} Charge-charge structure factors $S_\text{CC}(k)$
  for selected temperatures (as indicated) along the isochore $\rho=0.01$ from MD
  simulations (filled symbols) and from Eq.~\eqref{eqn:skrpa} (solid lines). The dashed line indicates a
  quadratic fit $(k/\kappa)^2$, where $\kappa$ is an effective inverse screening length, to the low-$k$ behaviour of
  $S_\text{CC}(k)$ for $T=0.025$.}
\end{figure}

From the Fourier transforms of the pair correlation functions
$h_{\alpha \beta}(r) = g_{\alpha \beta}(r) -1$ ($\alpha, \beta = $ +
or -), we extract the charge-charge structure factor
\begin{eqnarray} \nonumber
S_\text{CC}(k) & = & 1 + n [ \hat h_{++}(k) - \hat h_{+-}(k) ] \\ 
\label{eqn:skrpa}
& \approx & \frac{k^2}{k^2 + \kappa_{\rm D}^2 \exp[-k^2 \sigma^2]} ~~~~ {\rm (RPA)} 
\end{eqnarray}
where $\kappa_{\rm D} = \left( 8 \pi n Q^2/\epsilon ' k_{\rm B} T
\right)^{1/2}$ is the inverse Debye screening length, $k$ is the
wave-number and hats imply Fourier transforms. The RPA incorporates the exact
Stillinger-Lovett sum rule~\cite{stillinger_ion-pair_1968} $\lim_{k
  \to 0} \kappa_{\rm D}^2 S_\text{CC}(k)/k^2 = 1$, valid for a conducting
medium. The sum rule is well satisfied by our simulation data for $S_\text{CC}(k)$
along the isochore $\rho = 1$ at all temperatures, and the RPA
expression [Eq.~\eqref{eqn:skrpa}] provides an accurate representation of the data for
all $k$ at sufficiently high $T$ (not shown here). Along the low density isochore $\rho = 0.01$ (see Fig.~\ref{fig:5}), the
RPA expression is accurate only for $T \gtrsim 0.5$ while
for $T \lesssim 0.1$, strong deviations from the perfect
screening sum rule are observed; fits to the small-$k$ parabolic
behaviour of $S_\text{CC}(k)$ lead to an effective inverse screening length
$\kappa$ systematically larger than $\kappa_{\rm D}$ (by a factor of
two at $T = 0.025$). This implies that the URPM no longer
behaves as a conductor but rather as a dielectric medium of neutral
anion/cation clusters.

\begin{figure}[tbp]
\includegraphics[width=\onefig]{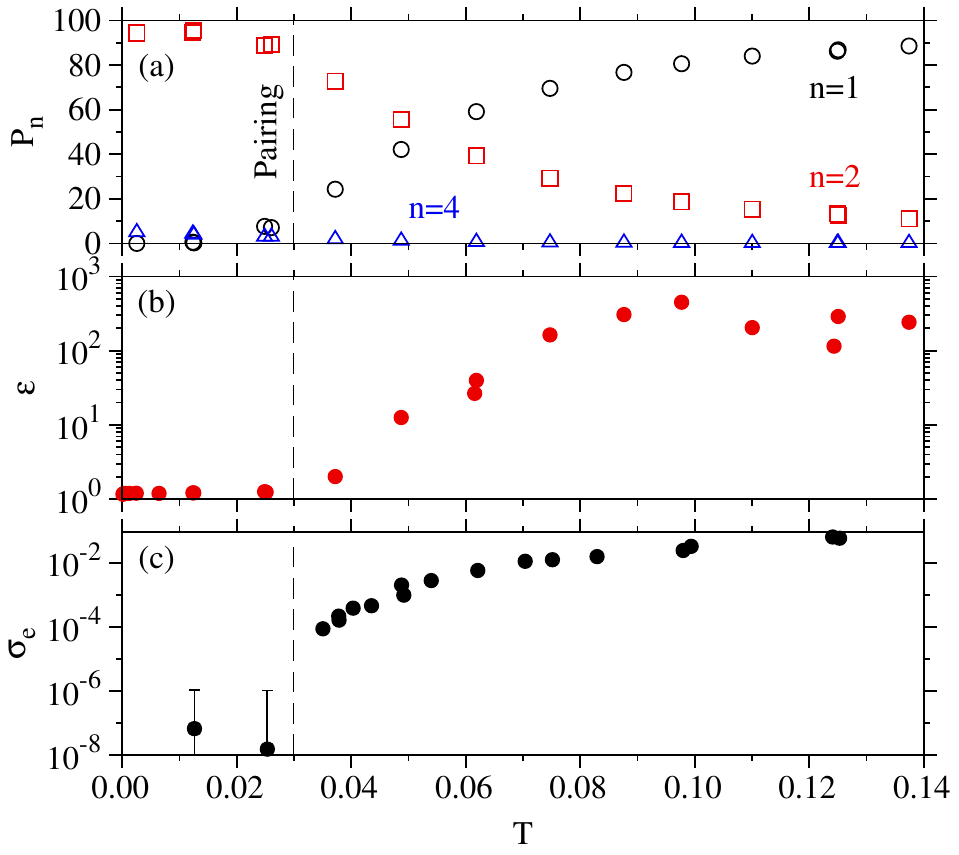}
\caption{\label{fig:2} Comparison of (a) percentages of selected $n$-mers $P_n$, (b) dielectric permittivity $\epsilon$, and
  (c) conductivity $\sigma_\text{e}$ as functions of $T$ along the isochore
  $\rho=0.01$.
}
\end{figure}

Using a standard geometric definition of $n$-ion clusters, namely that
$n$ ions form an $n$-mer if each ion lies within a distance $r_c$ ($=
1.0$) of at least one other ion in the cluster, we have determined the percentages $P_n$ of $n$-mers,
averaged over all configurations. Examples of $P_1$ (isolated ions),
$P_2$ (anion/cation pairs) and $P_4$ (tetramers) as functions of $T$ are
shown in Fig.~\ref{fig:2}-a, along the isochore $\rho =
0.01$. As expected, $P_2$ is close to 100\% at the lowest $T$,
and drops rapidly for $T \gtrsim 0.03$, while $P_1$ starts from
0 and increases towards 100\% for $T \gtrsim 0.03$. Note that
the percentage of trimers (now shown) is always negligible
while the percentage of tetramers can be significant ($\agt 5$\% at the lowest temperatures). The lifetime $\tau_2$ of pairs, as estimated from
MD simulations, increases dramatically as $T$ drops below 0.05, and is approximately 
two orders of magnitude larger than that of the other $n$-mers.

To assess explicitly the effect of pairing on the dielectric properties of the system,
we calculate the
dielectric permittivity $\epsilon$ from the
fluctuations of the total dipole moment of the simulated periodic
sample, ${\bf M} = \sum_i Q_i {\bf r}_i$,
according to the standard Kirkwood
relation~\cite{kirkwood_dielectric_1939}. Results from our
simulations along the isochore $\rho = 0.01$ are shown in
Fig.~\ref{fig:2}-b. For $T \lesssim 0.03$ the fluctuations of
${\bf M}$ are small, and the resulting $\epsilon$ takes on values of
the order of $1.3$, typical of dielectrics made up of polarizable
molecules (ion pairs in our case). At higher $T$ fluctuations of
${\bf M}$ are strongly enhanced, due to the break-up of ion pairs, and
$\epsilon$ rises sharply towards values larger than 100, typical of a
conductor (for an infinite conducting sample, $\epsilon \to \infty$).

\begin{figure}[t]
\includegraphics[width=\onefig]{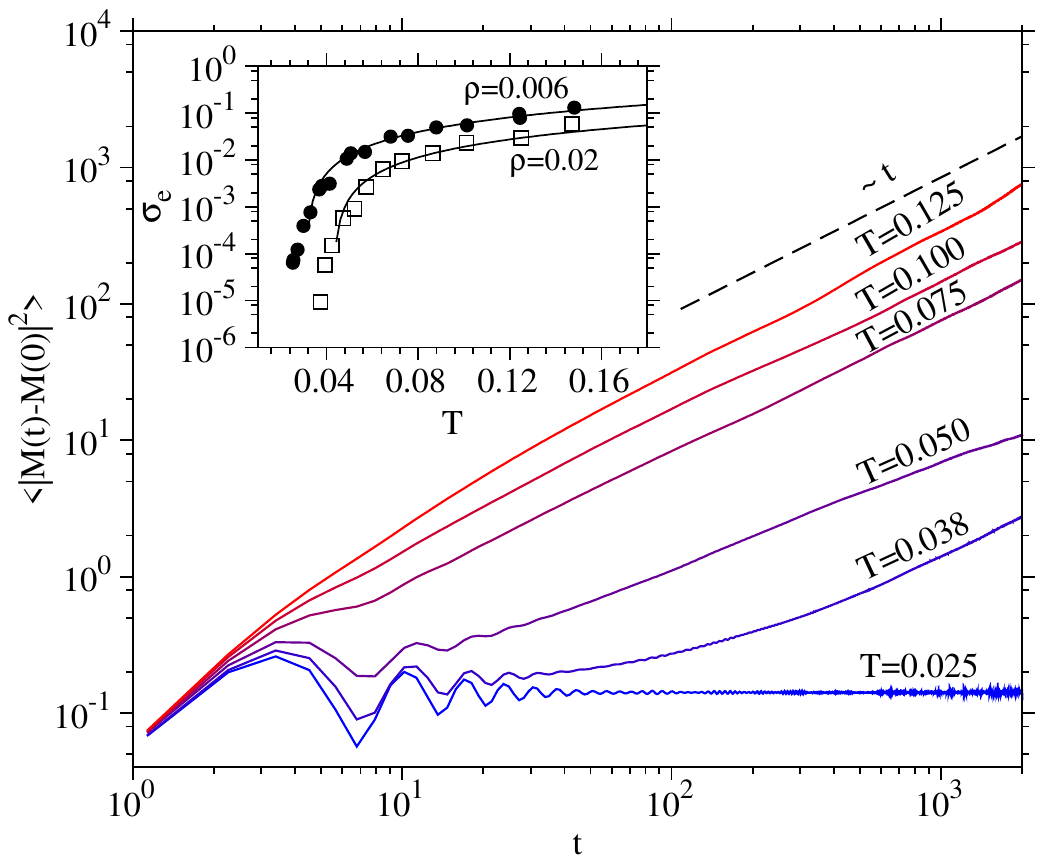}
\caption{\label{fig:3} Diffusion of the
  total dipole moment $c(t) = \langle | \ve{M}(t) - \ve{M}(0) | ^2
  \rangle$ as a function of reduced time $t$ for selected temperatures (as indicated) along the
  isochore $\rho=0.01$.
  Inset:
  electrical conductivity $\sigma_\text{e}$ as a function of $T$ for
  $\rho=0.006$ (filled circles) and $\rho=0.02$ (empty
  squares). Solid lines are power law fits $\sigma_\text{e}(T) \approx
  A(T-T_\sigma)^{1.2}$.}
\end{figure}

These suggestions of a CI transition prompted
us to extract some dynamic diagnostics from MD simulations~\footnote{Brownian
Dynamics simulations would obviously be more appropriate to account
for solvent friction, but the more efficient MD, capable of exploring
longer time scales was preferred, because we are interested in
qualitative, rather than quantitative aspects of the
dynamics.}. Figure~\ref{fig:3} shows plots of the time-dependent
electric dipole diffusion $c(t) = \langle | {\bf M}(t) - {\bf M}(0) |
^2 \rangle$. For a conductor, $c(t)$ increases linearly at long times
(according to the generalized Einstein relation), and the asymptotic slope is
proportional to the electrical conductivity $\sigma_\text{e}$. At the lowest
$T$, the slope vanishes, i.e., $\sigma_\text{e} = 0$, corresponding to a
dielectric insulator state. The reduced angular frequency of the
oscillations observed at low $T$ is close to the harmonic oscillator frequency of the
parabolic anion/cation potential well [Eq.~\eqref{eqn:1b}], confirming
the formation of long-lived pairs that do not contribute to the
electrical conductivity. The temperature variation of $\sigma_\text{e}$
is illustrated in the inset of Fig.~\ref{fig:3} along two isochores
and confirms the sharp drop of $\sigma_\text{e}$ by several orders of
magnitude over a narrow range of temperatures. The conductivity data
have been fitted by power laws $A(T-T_\sigma)^{1.2}$, % assuming a fixed, common value $\gamma=1.2$,
yielding the apparent CI transition temperatures
$T_\sigma=T_\sigma (\rho)$.

Figure~\ref{fig:2} provides an overview of the behaviour of the system
along the representative isochore $\rho = 0.01$, displaying
results for the percentages $P_n$ of $n$-mers, the permittivity $\epsilon$, and the conductivity
$\sigma_\text{e}$ 
as functions of $T$. The three plots show a strong correlation
between the sharp rise in $\epsilon$, $\sigma_\text{e}$ and $P_1$ around $T
= 0.03$, indicative of a CI transition, driven by
pairing and rounded by finite size effects. 

\begin{figure}[tbp]
\includegraphics[width=\onefig]{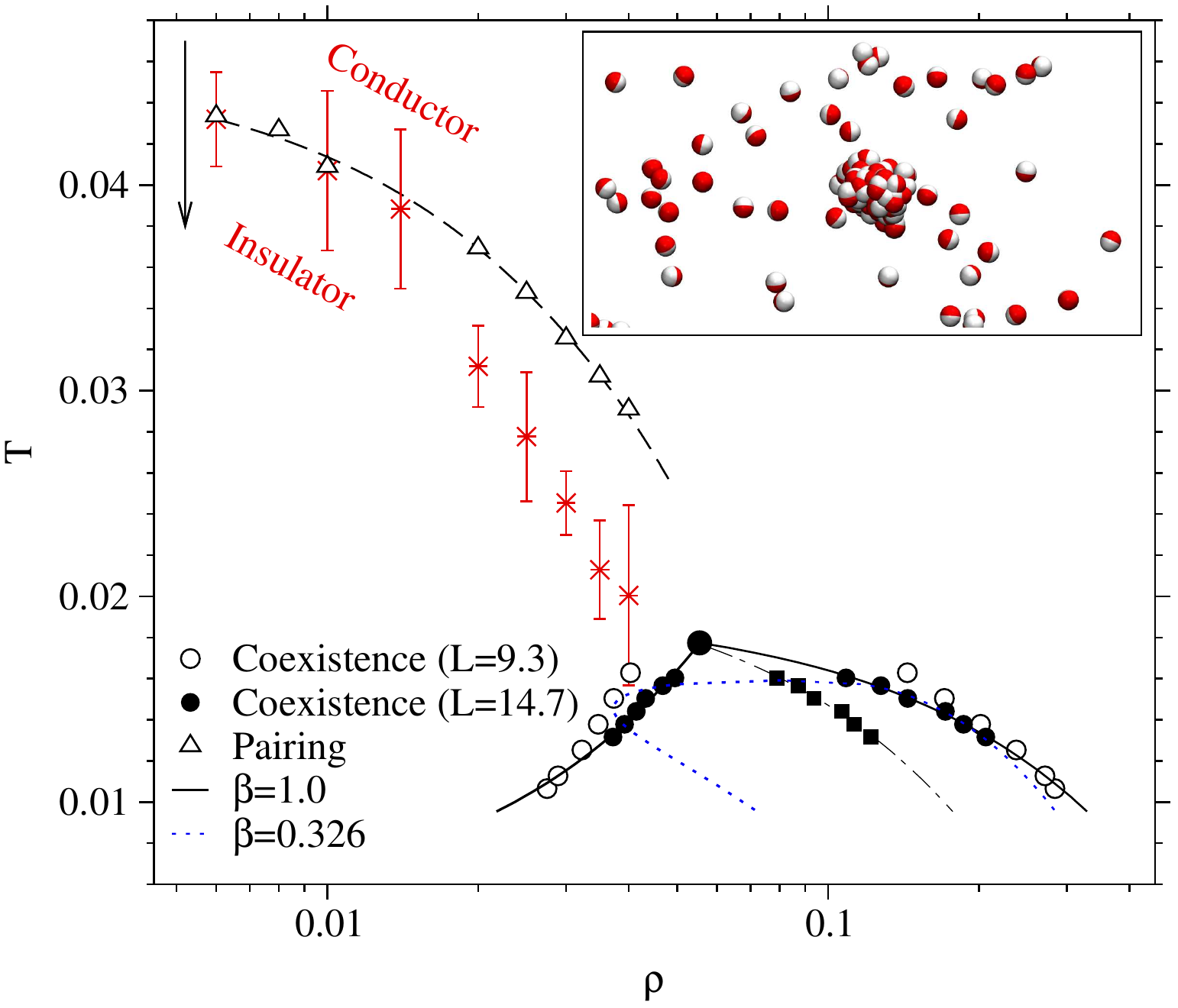}
\caption{\label{fig:4} Extended phase diagram of the URPM:
  liquid-vapor coexistence from GCMC simulations ($L=14.7$, filled
  circles; $L=9.3$, empty circles), pairing transition points $(T_p,
  \rho_p)$ determined from the condition $P_1(T_p,\rho_p)=30\%$ (empty
  triangles) and CI transition points $(T_\sigma, \rho_\sigma)$
  (stars; see text for definition). Critical fits to the coexistence curve for
  $L=14.7$ are shown for $\beta=0.326$ (dotted line) or $\beta=1.0$
  (solid line). The estimated critical point corresponding to 
  $\beta=1.0$ is shown as a bigger filled sphere. The
  law of rectlinear diameter is also included
  (dash-dotted line and filled squares). The dashed line through the pairing transition
  points is drawn as a guide to the eye. Inset: MD-generated
  configuration at $\rho=0.01$ and $T=0.00125$. White and red spheres
  (not drawn to scale) indicate oppositely charged particles.}
\end{figure}

The final step of our investigation is to relate the clustering to the
liquid-vapour phase transition expected at low $T$. To that purpose we
have carried out grand-canonical MC simulations in periodic cells of
side $L=9.28$ and $L=14.7$, combining biased MC sampling and histogram
reweighting techniques~\cite{wilding_computer_2001} to construct the
coexistence curves shown in Fig.~\ref{fig:4}, together with the loci
of state points where the CI transition is expected (i.e., $\sigma_\text{e}
\to 0$ according to the aforementioned power law fits) and where the
fraction $P_1$ of free ions reaches 30\%. Different choices of the value
of $P_1$ at the pairing transition temperature, as well as variations
of $r_c$ by some 10\% or 20\%, give rise to pairing transition temperatures
shifted by at most $\pm 0.01$ with respect to those displayed in
Fig.~\ref{fig:4}. The extrapolation of the two transition
lines intersect the coexistence curve close to the critical point,
roughly estimated to be $T_c \sim 0.018$, $\rho_c \sim 0.05$ on the
basis of the results for $L=14.7$. The phase diagram shows that the
liquid-vapour coexistence curve is better fitted by a scaling law with
a critical exponent $\beta=1$ than by the Ising universality class
exponent $\beta=0.326$ (or by the mean-field exponent $\beta=0.5$, not
shown). Inclusion of corrections to scaling did not improve
the fit for $\beta=0.326$ substantially. The break-down of an
Ising-like description of the coexistence curve, the apparent value of
the fitted critical exponent ($\beta\approx 1$) and the proximity of
the critical point to the CI line hints at a tricritical nature of the
phase diagram of the URPM, very different from the Ising-like phase
diagram of the RPM. This conjecture will be tested by finite size
scaling calculations in future work.  We note that the putative
tricritical behavior of our model may be different in nature from the
one observed in a lattice version of the RPM~(see~\cite{ren_2006} and
references therein). The latter behavior is linked in fact to an
order-disorder transition, reminiscent of that observed in some
antiferromagnetic materials.
% Note that tricritical behaviour has been
% observed in simulations of a discrete lattice version of the
% RPM~(see~\cite{ren_2006} and references therein), but that behavior is
% linked to an order-disorder transition, reminiscent of that observed
% in some antiferromagnetic materials; it seems thus unrelated to the
% behavior observed in the present, continuous model, where
% tricriticality is intimately related to pairing.

As in the case of the RPM, the dilute phase is dominated by long-lived
ion pairs (and larger neutral clusters), while in the dense phase,
ions are essentially free. This is illustrated by the snapshot in the
inset of Fig.~\ref{fig:4}, which shows droplet formation during an MD
simulation at low density and temperature.  Identifying the length
scales $\sigma$ of the RPM and the URPM, the reduced critical
parameters $T_c$ and $\rho_c$ of the two models are of the same order
of magnitude, suggesting that the phase separation is essentially of
Coulombic origin. However, a closer comparison between
Fig.~\ref{fig:4} and recent simulation results for the
RPM~\cite{valeriani_ion_2010} reveals a striking difference in the
topology of the phase diagram: while in our model the pairing
transition line intersects the coexistence curve close to the estimated
critical point, pair formation in the RPM represents a crossover
occurring at densities much lower than the critical one. Thus the
discrepancies between the hard core and soft core models may be traced
back to the fundamental difference between long-lived ion pairs at low
temperatures and densities: these pairs are strongly dipolar dumbbells
in the RPM, while in the present URPM they are non-polar, polarizable
entities.

In summary, we have introduced a simple model of oppositely charged,
interpenetrating macroions, which generalizes the familiar RPM to
``soft'' polyelectrolytes. Using static and dynamic diagnostics in MC
oand MD simulations, we have provided a quantitative characterization
of pairing and clustering of ions at low temperatures and densities,
and their impact on the segregation into coexisting dilute and
concentrated phases. The predicted clustering and segregation of the URPM are reminiscent of the
experimentally observed complexation of anionic and cationic
polyelectrolytes and subsequent complex coacervation~\cite{phillip_98,buchhammer_03}, as confirmed
by approximate field-theoretic calculations and simulations~\cite{castelnovo_01,lee_08}. Examination of the state-dependence of the
dielectric pemittivity and of the electric conductivity suggests that
pairing leads to a CI transition resembling that observed in
liquid metals, such as Hg or Rb. 
Future work will concentrate on a quantitative analysis of
finite size effects and the extension of the URPM to the unrestricted,
asymmetric case.

\begin{acknowledgements} 
We thank C. Valeriani, K. Binder, and S. Clarke for useful discussions.
D.C. and G.K. acknowledge financial support by the Austrian
Science Foundation (FWF) under Proj. No. P19890-N16.
\end{acknowledgements}

\providecommand*{\mcitethebibliography}{\thebibliography}
\csname @ifundefined\endcsname{endmcitethebibliography}
{\let\endmcitethebibliography\endthebibliography}{}

% \bibliographystyle{rsc}
% \bibliography{biblio}

\end{document}